# Beam optical design for high energy electron radiography experiment study based on THU LINAC

Quantang Zhao(赵全堂)[1,#], Shuchun Cao(曹树春)[1], Ming Liu(刘铭)[1], Xiaokan Sheng(申晓康)[1], Yanru Wang(王燕茹)[1,2], Yang Zong(宗阳)[1], Yi Jing(景漪)[1], Xiaoming Zhang(张校铭)[1,2], Rui Cheng(程锐)[1], Yongtao Zhao(赵永涛)[1], Zimin Zhang(张子民)[1,#], Yingchao Du(杜应超)[3], Wei Gai (盖炜)[4]

[1]Institute of Modern Physics, Chinese Academy of Sciences, Lanzhou 730000, China

[2]University of Chinese Academy of Sciences, Beijing 100049, China

[3]Department of engineering physics, TsingHua University, Beijing 100084, China

[4]Argonne national laboratory, Argonne, IL 60439, USA

Abstract

　　A special beam line for high energy electron radiography is designed, including achromat and imaging systems. The requirement of the angle and position correction on the target from imaging system can be approximately realized by fine tuning the quadrupoles used in the achromat. The imaging system is designed by fully considering the limitation from the laboratory and beam diagnostics devices space. Two kinds of imaging system are designed and both show a good performance of imaging by beam trajectory simulation. The details of the beam optical requirement and optimization design are presented here. The beam line is designed and prepared to install in Tsinghua university linear electron accelerator laboratory for further precise electron radiography experiment study.

## 1 Introduction

　　A new scheme is proposed that high energy electron beam as a probe used for time resolved imaging measurement of high energy density materials, especially for high energy density matter and inertial confinement fusion (ICF) [1-2]. High Energy Density Physics aims to study the properties of matters under extreme temperature and pressure state, which is also called Warm Dense Matter (WDM). According to WDM properties, the high energy density exceeds 1 Mbar and the transiently produced in laboratory experiment on the 10 ns to 1 us time scale. The diagnostics system should have a large dynamic range and high spatial resolution. Furthermore, for ICF, it is essential to measure the moving boundary, so the time dependent imaging system is desirable. Comparing with proton and some other x-ray diagnostics system, electron imaging system is expected to gain high spatial and temporal resolution with less expense. The high electron radiography (eRad) is developed by Los Alamos National Lab [3]. The first picosecond pulse-width high electron radiography experiment was achieved by Institute of Modern Physics (IMP), Chinese Academy of Sciences (CAS) and Tsinghua University (THU), based on THU Linear electron accelerator (LINAC) [4]. It is used for principle test and certifying that this kind of

*Supported by the National Natural Science Foundation of China11435015, 11505251 and 10921504.

# Email: zzm@impcas.ac.cn

# Email: zhaoquantang@impcas.ac.cn

LINAC with ultra-short pulse electron bunch can be used for electron radiography. Although the imaging system is not optimized, the experiment results, such as magnifying factor and the imaging distortion, are consistent with the beam optical theory very well.

For further experiment study, a new special beam line for eRad experiment is designed and planned to be constructed in THU LINAC laboratory. The LINAC [5] consists of s-band photocathode microwave electron gun and an s-band accelerating tube. The photocathode microwave electron gun provides low emittance and picosecond pulse-width electron bunch with energy of 3 MeV. The highest beam energy of the LINAC is 50 MeV. For the eRad experiment, the energy of 40 MeV is chosen as requested for the LINAC stable running. The beam properties of this LINAC used in the eRad are as follows: beam spot size radius is about 1.5 mm, divergence angle is about 1.5 mrad and the energy spread is less 1%.

## 2 Beam optical layout and design

Due to some reasons, such as not interrupt other experiment, a beam exacted from the linear beam line of the LINAC is used and specially designed for the eRad experiment. The beam line design is mainly limited by the small space of the laboratory, so the first consideration is the layout of the beam line. According to the laboratory space, the layout is shown in Fig.1. A linear achromat consist of two rectangular dipoles and three quadrupoles are used for deflecting the beam to 90 degree. The imaging target is placed after the achromat in the object plane position. After the target, it is the imaging system. For more detailed imaging system optical property study, two kinds of imaging system are designed with different magnification factor, with symmetric doublet and triplet quadrupoles. Also the aperture is placed on the Fourier plane position.

We first give the design parameters of the imaging system and then give the achromat design, because there are some special requirements of the achromat from the imaging system, such as beam angle and position calibration on the target.

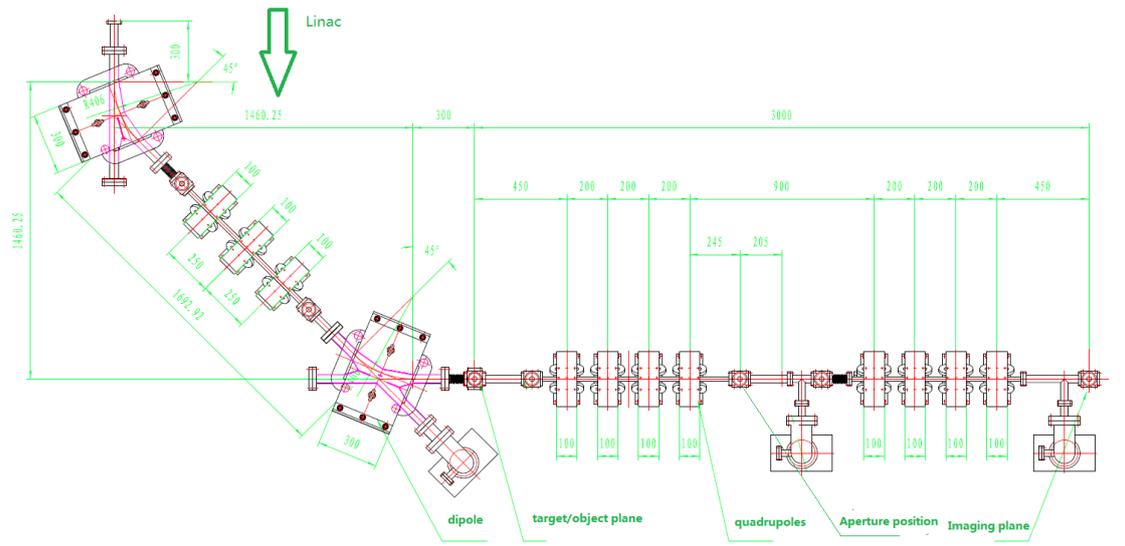

Fig.1. the layout of the eRad beam line based on THU linac

## 2.1 Beam optical design for eRad

There are two primary requirements of any charged particle radiography (CPR) lens system

[6]. First, the lens must provide a point-to-point focus from object to image. Second, it must form a Fourier plane, where particles are radially sorted by the magnitude of the scattering within the object. With the correlation, particles which were scattered to large angles by multiple Coulomb scattering can be removed through collimation at the Fourier plane. The remaining parameters of one experiment setup CPR lens system design are determined by the radiographic applications. The beam energy must high enough to penetrate the areal density of the object to be radiographed, and the aperture of the lens system must be chosen to provide sufficient angular acceptance throughout the required field of view. An additional strongest design requirement is the resolution of the radiography system. This resolution is typically dominated by chromatic aberrations due to energy spread of the injected beam in combination with the spread of energy loss through the object due to areal density variations of the object.

Here present some requirements and conditions for the charged particle radiography from beam optical with transport matrix. For the first order beam optical design, the beam on image plane can be calculated by the transport matrix as follows [7],

$$x_{\text{image}} = R_{11}x + R_{12}\theta$$

$$y_{image} = R_{33}y + R_{34}\psi \tag{1}$$

Considering the x direction firstly, for the first order beam optical, the beam position after the lens is

$$x_{fir\_image} = R_{11}x + R_{12}\theta \tag{2}$$

For the second order transport matrix, the beam on image plane is

$$x_{sec\_image} = R_{11}x + R_{12}\theta + T_{116}x\delta + T_{126}\theta\delta \tag{3}$$

R is the first order transport matrix, T is the second order transport matrix (Transport notation [8]), beam position is $(x, y)$, the divergence angle is $(\theta, \psi)$, and $\delta$ is the beam momentum spread.

The conditions for point to point imaging is required $R_{12}=0$ and $R_{11}$ is the magnification factor. On the second order beam optical cases:

$$x_{sec\_image} = R_{11}x + T_{116}x\delta + T_{126}\theta\delta$$

$$x_{sec\_image}/R_{11} = x + (T_{116}x\delta + T_{126}\theta\delta)/R_{11} \tag{4}$$

The space resolution is defined as follows:

$$\Delta x = x - x_{sec\_image}/R_{11} = (T_{116}x\delta + T_{126}\theta\delta)/R_{11}$$

$$\Delta x = (T_{116}x\delta + T_{126}\theta\delta)/M_x \tag{5}$$

With same procedure, the resolution on y direction is as follows,

$$\Delta y = (T_{336}y\delta + T_{346}\psi\delta)/M_y \tag{6}$$

Assuming $\theta = wx$, the residual incident emittance angles are added to the multiple coulomb scattering angles produced by the object, so the relationship of position and angle can be corrected as $\theta = wx + \varphi$, define $w = \dfrac{\theta}{x} = -\dfrac{T_{116}}{T_{126}}$, so

$$\theta = -\dfrac{T_{116}}{T_{126}} x + \varphi \qquad (7)$$

$\varphi$ is the beam scattering angle after interacting with the target with residual incident angles.

After the above calibration of beam angle on the target, the space resolution can be defined as follows, which depends on the chromatic length, scattering angle with residual incident angles and beam momentum spread.

$$\Delta x = L_c \varphi \delta \qquad (8)$$

$L_c = \dfrac{T_{126}}{M}$ is the chromatic length

Normally, there is an aperture on the Fourier plane, which limits the beam with small divergence angle $\varepsilon$ ($\epsilon < \varphi$) can pass through. So the spatial resolution with aperture is defined:

$$\Delta x = L_c \varphi \varepsilon \qquad (9)$$

2.1.1 Doublet imaging system design

As mentioned before, the imaging system is limited by the space. The doublet design parameters are optimized by COSY INFINITY9.1 [9] by only tuning the quadrupoles field. The maximum filed gradient of the quadrupoles are 12 T/m, the inner diameter is 20 mm and the length is 10 cm. the imaging system optical with optimized quadrupoles field gradient is show in Fig.2, beam on object plane with different position and different divergence angle are perfectly imaged on the image plane. It has a magnification factor for $R_{11}=R_{33}=-2.885$. The detailed beam line parameters are shown in table1.

Table1 the optimized parameters of drifts and quadrupoles for doublet imaging system

| Quads | Flux density at pole tip B [T] | Transport matrix element | Optimized values | Drift | Drfit distance(m) |
|---|---|---|---|---|---|
| $Q_1$ | 0.02497 | $R_{11}=R_{33}$ | -2.8847 | $L_1$ | 0.4 |
| $Q_2$ | -0.06113 | $R_{12}=R_{34}$ [m/rad] | 0 | $L_2$ | 0.1 |
| $Q_3$ | 0.06113 | $T_{116}$, $T_{126}$ [m/rad] | 6.8925, 4.8097 | $L_3$ | 0.1 |
| $Q_4$ | -0.02497 | $T_{336}$, $T_{346}$ [m/rad] | 6.088, 6.0164 | $L_4$ | 1.9 |

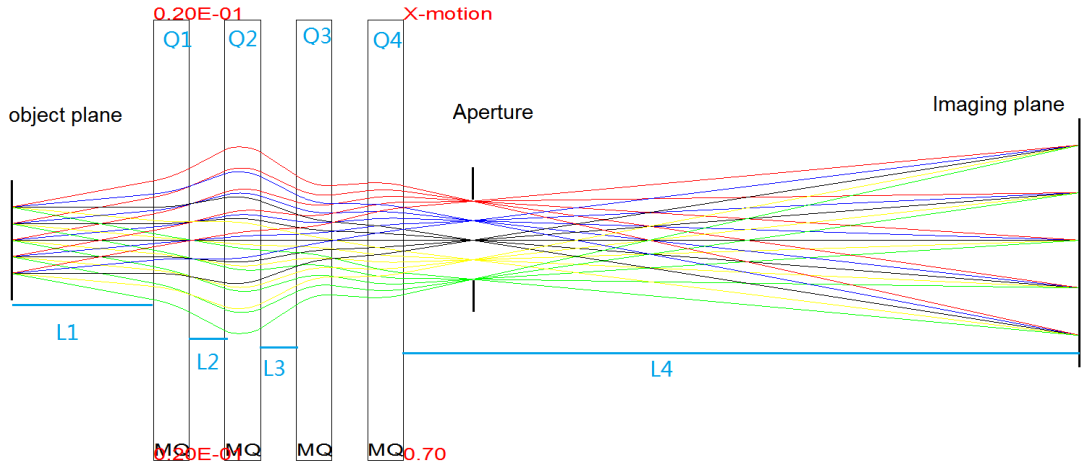

Fig.2. the beam trajectory of the doublet imaging system in X direction.

### 2.1.2 Triplet imaging system design

Another imaging system is consisting of two triplet quadrupoles. All the quadruploes are the same design for convenient production. The beam optical is shown in Fig.3 with optimized field gradient. It has a unit magnification factor, $R_{11}=R_{33}=-1$. It also shows a good performance of the imaging property. The detailed beam line parameters are shown in table2.

Table2 the optimized parameters of drifts and quadrupoles for triplet imaging system

| Quads | Flux density at pole tip B [T] | Transport matrix element | Optimized values | Drift | Drift distance |
|---|---|---|---|---|---|
| $Q_1$ | 0.02945 | $R_{11}=R_{33}$ | -1.0 | $L_1$ | 0.4 |
| $Q_2$ | -0.02722 | $R_{12}=R_{34}$ [m/rad] | 0 | $L_2$ | 0.1 |
| $Q_3$ | -0.02722 | $T_{116}, T_{126}$ [m/rad] | 0, 3.0464 | $L_3$ | 0.8 |
| $Q_4$ | 0.02945 | $T_{336}, T_{346}$ [m/rad] | 0, 3.2409 | $L_4$ | 0.4 |

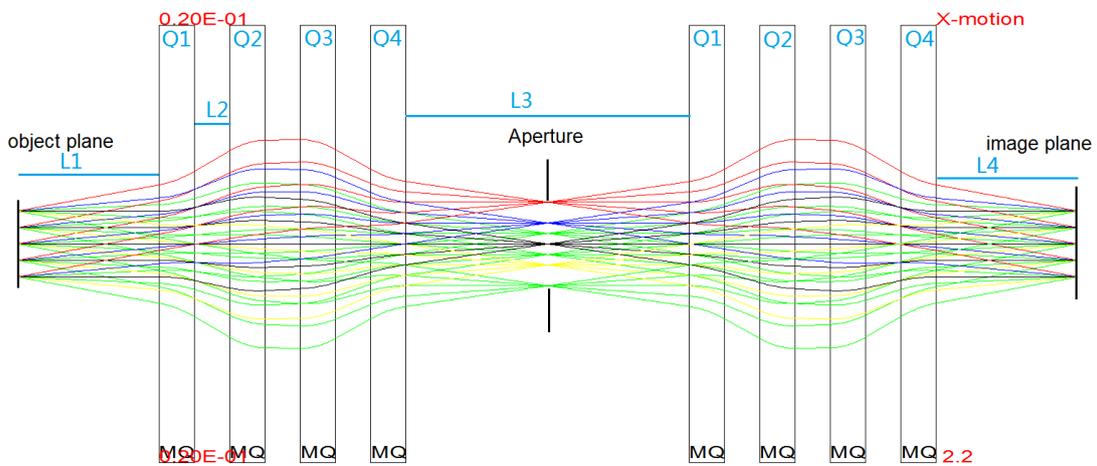

Fig.3. the beam trajectory of the triplet imaging system in X direction.

## 2.2 Linear achromat and beam angle and position calibration on the target

### 2.2.1 Linear achromat design

Usually it is necessary in beam transport systems to deflect a particle beam. If this is done in an arbitrary way an undesirable finite dispersion function will remain at the end of the deflecting section. Special magnet arrangement is designed which allow to bend a beam without generating a residual dispersion. Such magnet systems composed of only bending magnets and quadrupoles are called linear achromat.

Dispersion function is defined as follows with transport matrix elements [10]:

$$D(z) = S(z) \int_0^z k_0(z) C(\tilde{z}) d\tilde{z} - C(z) \int_0^z k_0(\tilde{z}) S(\tilde{z}) d\tilde{z}$$

The physical interpretation of the dispersion function D(z) is simply that the function δD(z) determines the offset of the reference trajectory from the ideal path for particles with a relative energy deviation δ from the ideal momentum $cP_0$.

The dispersion function generated in a particular bending magnet does not depend on the dispersion at the entrance to the bending magnet which may have been generated by upstream bending magnets. The dispersion generated by a particular bending magnet reaches the value $D(L_m)$ at the exit of the bending magnet of length $L_m$ and propagates from there on through the rest of the beam line just like any other particle trajectory, Which has exactly the form of describing the trajectory of a particle starting with initial parameters at the end of the bending magnet given by the integrals. With this solution we can expand the 2×2 matrix in to 3×3 matrix, which includes the first order chromatic correction.

$$\begin{bmatrix} u(z) \\ u'(z) \\ \delta \end{bmatrix} = \begin{bmatrix} C_u(z) & S_u(z) & D_u(z) \\ C'_u(z) & S'_u(z) & D'_u(z) \\ 0 & 0 & 1 \end{bmatrix} \begin{bmatrix} u(z_0) \\ u'(z_0) \\ \delta \end{bmatrix}$$

The achromat requires both the dispersion and its derivative to vanish, $D(z_d) = 0$ and $D'(z_d) = 0$, which corresponds to the $R_{16} = R_{26} = 0$ in x direction. This can be done with some special settings for the lattice and adjusting the quadrupole fields. In this case we have no dispersion function downstream the point $z = z_d$ up to the point where the next dipole magnet creates a new dispersion function. The physical characteristics of an achromatic beam line are that at the end of the beam line, the position and the slope of a particle are independent of the energy.

For our achromat, two rectangular dipole magnets are used for bending the beam. They are installed symmetrically for the intended particle trajectory. The entrance and exit angles equal half of the bending angle -22.5 degree. The rectangular magnet has edge focusing properties in the non-bending plane. In the deflecting plane the focusing is completely eliminated.

In COSY INFINITY9.1 [9], a special case of the homogeneous dipole described above is the magnetic rectangle or parallel-faced dipole, in which both edge angles equal one half of the deflection angle and the curvatures are zero. The parameters of rectangle dipole used in current experiment are as follows: the radius is 392 mm, deflecting angle is 45 degree and the aperture is 25 mm.

### 2.2.2 Achromat with angle position correlation on target design

As derived on section 2.1, there are some requirements for beam angle and position calibration from imaging system. The angle position correlation is derived from the imaging system, its second order beam optical matrix T [7].

$$W_x = -\frac{T_{116}}{T_{126}}, \quad W_y = -\frac{T_{336}}{T_{346}}$$

It means the beam on target have a correlation between position and incident angle. Based on doublet imaging system the angle-position correlation is as follows:

$$w_x = -T_{116}/T_{126} = -1.433, \quad w_y = -T_{336}/T_{346} = -1.0119$$

So these constrains are added to the achromat optimization, try to find the solutions with tuning the quadrupoles fields. For doublet imaging system, the optimized achromat with certain angle and position correction beam optics and trajectory are shown in Fig. 4(a) and Fig. 4(b). It is shown that for the achromat, it is also can approximately give the angle position correction required by the imaging system. Another requirement of the beam on the target is the beam spot size, which should be large enough to illuminate the object area of interesting. The beam size is also shown in table 3 by beam matrix element, and the diameter of the beam spot size is about 3 mm. The parameters of the achromat design are shown in table 3. For the triplet imaging system design, $T_{116}$ nearly equals to zero, so there needs nearly the parallel beam.

Table3 the optimized parameters of drifts and quadrupoles for achromat

| Quads | Flux density at pole tip B [T] | Transport matrix and beam matrix element | Optimized values | Drift | Drift distance(m) |
|---|---|---|---|---|---|
| $Q_1$ | -0.04966 | $R_{16}$ [m] | -0.19E-9 | $L_1$ | 0.3 |
| $Q_2$ | 0.06137 | $R_{26}$ [rad] | 0.62E-10 | $L_2$ | 0.2 |
| $Q_3$ | -0.04966 | $sqrt(\sigma_{11}), sqrt(\sigma_{33})$ [mm] | 1.8, 1.6 | $L_3$ | 0.3 |

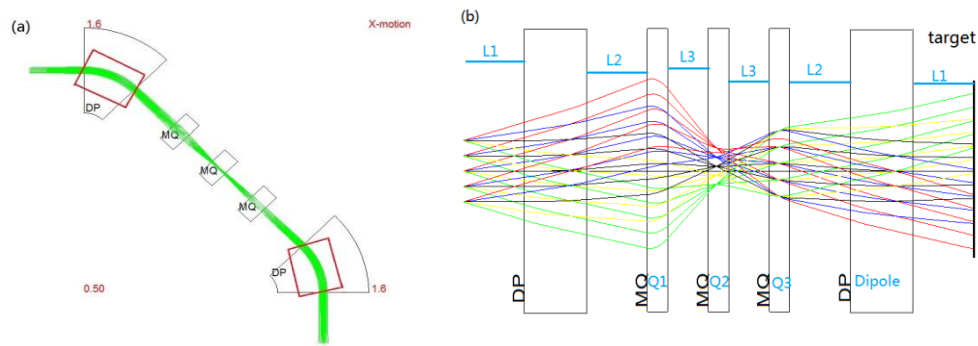

Fig.3. Achromat wiht beam angle and position calibration on the target in x direction according to doublet imaging system optical (a) and beam trajectory (b).

## 3 conclusions

A special beam line for high energy electron radiography experiment study with an achromat consisting of two dipoles and three quadrupoles is designed based on THU LINAC. It is confirmed that from the achromat, the requirement of the angle and position correction on the target from imaging system can be approximately realized by fine tuning the quadrupoles used in the achromat. The imaging system is designed by fully considering the laboratory and beam diagnostics devices space. Two kinds of imaging system are designed and both show a good performance of imaging by beam trajectory trace simulation. This is helpful for the precisely eRad experiment, such as imaging optical study and experiment results analyses. The beam line will be built and installed at THU LINAC laboratory and scheduled static and dynamic eRad experiments will be taken on this beam line.

# Acknowledgements

The work is supported by National Natural Science Foundation of China 11435015, 11505251 and 10921504.

# Reference

1 Y. T. Zhao, Z. M. Zhang, H. S. Xu, W. L. Zhan et al. A HIGH RESOLUTION SPATIAL-TEMPORAL IMAGING DIAGNOSTIC FOR HIGH ENERGY DENSITY PHYSICS EXPERIMENTS Proceedings of IPAC2014, Dresden, Germany, THOAB03

2 Wei Gai, Jiaqi Qiu, Chunuang Jing. Electron imaging system for untrafast diagnostics of HEDLP. Proc. of SPIE Vol. 9211, 921104.

3 Frank Merrill, Frank Harmon, Alan Hunt, et al. Electron radiography, Nuclear Instruments and Methods in Physics Research B 261 (2007) 382–386.

4 Quantang Zhao, Shuchun Cao, Rui Cheng, et al. HIGH ENERGY ELECTRON RADIOGRAPHY EXPERIMENT RESEARCH BASED ON PICOSENCOND PULSE WIDTH BUNCH. Proceedings of LINAC2014, Geneva, Switzerland, MOPP015.

5 Yingchao Du, Lixin Yan, Jianfei Hua, Qiang Du et al. Generation of first hard X-ray pulse at Tsinghua Thomson Scattering X-ray Source. Review of Scientific Instruments 84, 053301, 2013.

6 C L Morris, N S P King, K Kwiatkowski, F G Mariam, F E Merrill and A Saunders. Charged particle radiography, Rep. Prog. Phys. 76 (2013) 046301 (26pp).

7 C. T. Mottershead and J. D. Zumbro, in Proceedings of the Particle Accelerator Conference, Vancouver, Canada, 1997 (IEEE, Vancouver, 1997), p. 1397.

8 PSI Graphic Transport Framework by U. Rohrer based on a CERN-SLAC-FERMILAB version by K.L. Brown et al.

9 M. Berz and K. Makino. COSY INFINITY 9.1 Beam Physics Manual. MSU Report MSUHEP 060804-rev, August 2013.

10 Helmut Wiedemann. Particle Accelerator Physics, third edition. Springer, 2007.